\documentstyle[11pt,amssymb,amstex]{amsart}

\newcommand{\bea}{\begin{eqnarray}}
\newcommand{\eea}{\end{eqnarray}}
\newcommand{\ba}{\begin{array}}
\newcommand{\ea}{\end{array}}
\newcommand{\edc}{\end{document}}
\newcommand{\bc}{\begin{center}}
\newcommand{\ec}{\end{center}}
\newcommand{\be}{\begin{equation}}
\newcommand{\ee}{\end{equation}}

\newcommand{\dsf}{\displaystyle\frac}
\def\s{\sigma}
\def\e{\varepsilon}
\def\l{\lambda}

\def\t{\theta}

\def\a{\alpha}

\def\O{\Omega}

\def\R{\mathbb{R}}
\def\b{\beta}
\def\m{\mu}

\def\G{\Gamma}

\begin{document}

\title[on the Ising   model with competing interactions]
{on some remarks on the Ising model with competing interactions on
a cayley tree}

\author{Farrukh Mukhamedov}
\address{Farrukh Mukhemedov\\
Department of Mechanics and \\
Mathematics,National University \\
of Uzbekistan, Vuzgorodok,\\ 700095, Tashkent, Uzbekistan}
\email{{\tt far75m@@yandex.ru}}

\author{Utkir Rozikov}
\address{Utkir Rozikov\\
Institute of Mathematics,\\
 29, F. Hodjaev str., \\
 Tashkent, 700143,\\ Uzbekistan}
\email{{\tt rozikovu@@yandex.ru}}

\maketitle

\begin{abstract}
In the present paper the Ising  model with competing binary $J$ and
$J_1$ interactions with spin values $\pm 1$, on a Cayley
tree is considered. We study translation-invatiant  Gibbs measures and
corresponding free energies ones.
\end{abstract}

\section{Introduction}

Nowadays the investigations of statistical mechanics on
non-amenable graphs is a modern growing topics (\cite{L}). One of
the non-amenable graph is a Cayley tree. The Cayley tree is not a
realistic lattice, however, its amazing topology makes the exact
calculation of various quantities possible \cite{L}. It is
believed that several among its interesting thermal properties
could persist for regular lattices, for which the exact
calculation is far intractable. Here we mentions that in 90's a
lot of research papers were devoted to studying of the classical
the Ising model, with two spin values $\pm 1$, on such Cayley tree
(see \cite{Pr},
\cite{BG},\cite{BRSSZ},\cite{BRZ1},\cite{BRZ2},\cite{GR},\cite{R}).
In the present paper we extend some results of the paper
\cite{GPW}, where it has been investigated the Ising model with
competing interactions, with spin values $\pm 1$, on a Cayley
tree.  Note that such models was studied extensively (see Refs.
\cite{MTA},\cite{SC},\cite{Mo1},\cite{Mo2}) since the appearance
of the Vannimenus model (see Ref.\cite{V}), in which the physical
motivations for the urgency of study such models was presented. In
all of these works no exact solutions of the phase transition
problem were found, but some solutions for specific parameter
values were presented. On the other hand while studying such
models it was discovered  the appearance of nontrivial magnetic
orderings \cite{MR}.

In the paper using the methods of \cite{MR} we exactly solve a
phase transition problem for the model, namely,  we calculated
critical curve such that there is a phase transitions above it,
and a single Gibbs state is found elsewhere (cp.\cite{GPW}). We
also find ground states of the model. Besides, we will find a form
of the free energy of the model under consideration. This gives us
to study some asymptotics one.

\section{Preliminaries}

Recall that  the Cayley tree $\Gamma^k$ of order $ k\geq 1 $ is an
infinite tree, i.e., a graph without cycles, such that each vertex
of which lies on $ k+1 $ edges. Let $\Gamma^k=(V, \Lambda),$ where
$V$ is the set of vertices of $ \Gamma^k$, $\Lambda$ is the set of
edges of $ \Gamma^k$. The vertices $x$ and $y$ are called {\it
nearest neighbors}, which is denoted by $l=<x,y>$ if there exists
an edge connecting them. A collection of the pairs
$<x,x_1>,...,<x_{d-1},y>$ is called a {\it path} from $x$ to $y$.
Then the distance $d(x,y), x,y\in V$, on the Cayley tree, is the
length of the shortest path from $x$ to $y$.

For the fixed $x^0\in V$ we set

$$ W_n=\{x\in V| d(x,x^0)=n\}, \ \ \ V_n=\cup_{m=1}^n W_m, $$
$$ L_n=\{l=<x,y>\in L | x,y\in V_n\}. $$
Denote $|x|=d(x,x^0)$, $x\in V$.

Denote
$$
S(x)=\{y\in W_{n+1} :  d(x,y)=1 \}, \ \ x\in W_n.
$$
The defined set is called the set of {\it direct successors}.
Observe that any vertex $x\neq x^0$ has $k$ direct successors and
$x^0$ has $k+1$.

Two vertices $x,y\in V$ is called {\it one level next-nearest-neighboring
vertices} if
there is a vertex $z\in V$ such that  $x,y\in S(z)$ and they are denoted by
$>x,y<$. In this case the vertices $x,z,y$ are called {\it ternary} and denoted
by $<x,z,y>$.

{\bf Proposition 2.1.}\cite{G} {\it There exists a one-to-one
correspondence between the set $V$ of vertices of the Cayley tree
of order $k\geq 1$ and the group $G_{k}$ of the free products of
$k+1$ cyclic groups  of the second order with generators
$a_1,a_2,...,a_{k+1}$.} \\

Consider a   left (resp. right) transformation shift on $G_{k}$
defined as: for $ g_0\in G_{k}$  we  put
$$
T_{g_0}h=g_0h \ \ (\textrm{resp.}\ \  T_{g_0}h=hg_0,) \ \  \forall
h\in G_{k}.
$$
It is easy  to see that  the set  of all left  (resp. right) shifts on
$G_{k}$ is  isomorphic to the group $G_{k}$.

We consider models where the spin takes values in the set
$\Phi=\{-1,1\}$ . A configuration $\s$ on $V$ is then defined as a function
 $x\in V\to\s(x)\in\Phi$; the set of all configurations coincides with
$\Omega=\Phi^{V}$. The Hamiltonian  of the Ising model with competing
interactions has the form
$$
H(\s)=-J \sum\limits_{>x,y<}{\s(x)\s(y)}
-J_1 \sum\limits_{<x,y>}{\s(x)\s(y)}  \eqno (2.1)
$$
where $J,J_1\in {\R}$ are coupling constants and
$\s\in \Omega$.

As usually one can define Gibbs measures of the model under
consideration (see, for example, \cite{S},\cite{MR}).

\section{On Gibbs measures}

In this section we give the construction of a special class of
limiting Gibbs measures for the  Ising model on a Cayley tree with competing
interactions.

Let $h:x\to {\R}$ be a real
valued function of $x\in V$. Given $n=1,2,...$ consider the probability
measure $\m^{(n)}$ on $\Phi^{V_n}$ defined by
$$
\mu^{(n)}(\s_n)=Z^{-1}_{n}\exp\{-\b H(\s_n)+\sum_{x\in W_n}
h_x\s(x)\}, \eqno(3.1)
$$
where
$$ H(\s_n)=-J
\sum\limits_{>x,y<: x,y\in V_n}{\s_n(x)\s_n(y)} -J_1
\sum\limits_{<x,y>: x,y\in V_n}{\s_n(x)\s_n(y)}, $$ and
$\b=\frac{1}{T}$ and  $\s_n:x\in V_n\to\s_n(x)$ and $Z_n$ is the
corresponding partition function:
$$
Z_n\equiv Z_n(\b,h)=\sum_{\tilde\s_n\in\Omega_{V_n}}\exp
\{-\b H(\tilde\s_n)+\sum_{x\in W_n}h_x\tilde\s(x)\}.
$$

The consistency condition for $\m^{(n)}(\s_n), n\geq 1$ is
$$
\sum_{\s^{(n)}}\m^{(n)}(\s_{n-1},\s^{(n)})=\m^{(n-1)}(\s_{n-1}),
\eqno(3.2)
$$
where $\s^{(n)}=\{\s(x), x\in W_n\}$.

Let $V_1\subset V_2\subset...$ $\cup_{n=1}^{\infty}V_n=V$ and
$\m_1,\m_2,...$ be a sequence of probability measures on
$\Phi^{V_1},\Phi^{V_2},...$ satisfying the consistency condition
(3.2). Then, according to the Kolmogorov theorem, (see, e.g. Ref.
\cite{Sh}) there is a unique limit Gibbs measure $\m_h$ on $\O$
such that for every $n=1,2,...$ and $\s_n\in\Phi^{V_n}$ the
following equality holds $$
\m\bigg(\{\s|_{V_n}=\s_n\}\bigg)=\m^{(n)}(\s_n). \eqno(3.3) $$

The following statement describes conditions on $h_x$ guaranteeing
the consistency condition of measures $\m^{(n)}(\s_n)$. In the sequel
for the simplicity we will consider the case $k=2$.

{\bf Theorem 3.1.}{\it  The measures  $\m^{(n)}(\s_n)$,
$n=1,2,...$ satisfy the consistency condition (3.2) if and only if
for any $x\in V$ the following equation holds: $$ h_x= {1 \over
2}\log\bigg(\frac{\theta_1^2\theta
e^{2(h_y+h_z)}+\theta_1(e^{2h_y}+ e^{2h_z})+\theta}{\theta
e^{2(h_y+h_z)}+\theta_1(e^{2h_y}+
e^{2h_z})+\theta_1^2\theta}\bigg) \eqno (3.4) $$ here
$\theta=e^{2\beta J}, \ \ \theta_1=e^{2\beta J_1}$ and $<y,x,z>$
are ternary neighbors.}

{\bf Proof.} {\it Necessity}. According to the consistency
condition (3.2) we have
$$ {Z_{n-1} \over Z_n}\sum_{\sigma^{(n)}}\exp
\{-\beta H_{n-1}(\sigma_{n-1})+$$
$$
\beta J_1\sum_{x\in W_{n-1},y,z\in S(x)}\sigma(x)(\sigma(y)+\sigma(z))+
\beta J\sum_{x\in W_{n-1},y,z\in S(x)}\sigma(y)\sigma(z)+
$$
$$
\sum_{x\in W_{n-1}}\sum_{y\in S(x)}h_y\sigma(y)\}=
\exp\{-\beta H_{n-1}(\sigma_{n-1})+\sum_{x\in W_{n-1}}h_x\sigma(x)\}.
\eqno (3.5)
$$
Whence we get
$$ {Z_{n-1} \over Z_n}\sum_{\sigma^{(n)}}\prod_{x\in W_{n-1}}
\exp\{\beta J_1\s(x)(\s(y)+\s(z))+\beta J\s(y)\s(z)+
$$
$$
h_y\sigma(y)+h_z\s(z)\}=
\prod_{x \in W_{n-1}}\exp\{h_x\sigma(x)\}.
\eqno (3.6)
$$

Let  $ x\in W_{n-1}$ and $S(x)=\{y,z\},\ \  \sigma_x^{(n)}=
\{\sigma(y),\s(z)\}. $ Then it is easy to see
that  $ \sigma^{(n)}=\cup_{x\in W_{n-1}} \sigma_x^{(n)}. $ Hence
$$ {Z_{n-1} \over Z_n}\prod_{x\in W_{n-1}}\sum_{\sigma_x^{(n)}}
\exp\{\beta J_1\s(x)(\s(y)+\s(z))+\beta J\s(y)\s(z)+
$$
$$
h_y\sigma(y)+h_z\s(z)\}=
\prod_{x \in W_{n-1}}\exp\{h_x\sigma(x)\}.
\eqno (3.7)
$$

Now fix $x\in W_{n-1}$ and rewrite (3.7) for the cases $ \sigma(x)=1 $ and
$ \sigma(x)=-1 $ then we can find
$$ {\sum_{\sigma_x^{(n)}=\{\s(y),\s(z)\}}
\exp\{\beta J_1(\s(y)+\s(z))+\beta J\s(y)\s(z)+h_y\sigma(y)+h_z\s(z)\}
\over \sum_{\sigma_x^{(n)}=\{\s(y),\s(z)\}}
\exp\{-\beta J_1(\s(y)+\s(z))+\beta J\s(y)\s(z)+
h_y\sigma(y)+h_z\s(z)\}}$$
$$
=\exp\{2h_x\}.
\eqno (3.8)
$$

Denote
$$
W_1=\exp(2J_1\beta+J\beta+h_y+h_z)+ \exp(-J\beta-h_y+h_z)+
$$
$$
\exp(-J\beta+h_y-h_z)+
\exp(-2J_1\beta+J\beta-h_y-h_z) $$
$$
W_{-1}=\exp(-2J_1\beta+J\beta+h_y+h_z)+
\exp(-J\beta-h_y+h_z)+$$
$$\exp(-J\beta+h_y-h_z)+
\exp(2J_1\beta+J\beta-h_y-h_z) .  $$
It then follows from (3.8) that
$$\exp\{2h_x\}={W_1 \over  W_{-1}}
\eqno (3.9)
$$
The equality  (3.9) implies (3.4).

{\it Sufficiency}.  Now assume that (3.4) is valid, then it
implies (3.9), and hence (3.8). From (3.8) we obtain
$$ \sum_{\sigma_x^{(n)}=\{\s(y),\s(z)\}}
\exp\{\beta J_1(\s(y)+\s(z))\s(x)+\beta J\s(y)\s(z)+h_y\sigma(y)+h_z\s(z)\}
=$$
$$
a(x)\exp\{\s(x)h_x\}, \eqno (3.10)$$
where $\s(x)=\pm 1.$
This  equality implies
$$\prod_{x\in W_{n-1}} \sum_{\sigma_x^{(n)}=\{\s(y),\s(z)\}}
\exp\{\beta J_1(\s(y)+\s(z))\s(x)+\beta J\s(y)\s(z)+h_y\sigma(y)+h_z\s(z)\}
=$$
$$\prod_{x\in W_{n-1}}a(x)\exp\{\s(x)h_x\}.\eqno(3.11)  $$
Writing $A_n=\prod_{x\in W_n}a(x) $ we find from (3.11)
$$Z_{n-1}A_{n-1}\mu^{(n-1)}(\sigma_{n-1})=
Z_n\sum_{\sigma^{(n)}}\mu_n(\sigma_{n-1},
\sigma^{(n)}). $$
Since each $\mu^{(n)},\ \  n\geq 1$ is a probability measure, we have
$$ \sum_{\s_{n-1}}\sum_{\s^{(n)}}\mu^{(n)}(\s_{n-1}, \s^{(n)})=1, \ \
\sum_{\s_{n-1}}\mu^{(n-1)}(\s_{n-1})=1.$$ Therefore from these
equalities we find
$$Z_{n-1}A_{n-1}=Z_n, \eqno(3.12)$$ which means that (3.2) holds.
This completes the proof.

According to  Theorem 3.1 the problem of describing of Gibbs
measures is reduced to the description of solutions of the
functional equation (3.4).

By  Proposition 2.1 any transformation $S$ of  the group $G_{k}$
induces a shift automorphism $\tilde  S: \O\to\O$ by
$$
(\tilde S\s)(h)=\s(Sh), \ \  h\in G_{k},\ \s\in \O.
$$

By ${\cal G}_{k}$ we  denote the set of all shifts of $\O$.

We say that a Gibbs measure $\m$ on $\O$ is {\it translation - invariant}
if for any $T\in {\cal G}_{k}$ the equality $\m(T(A))=\m(A)$ is
valid for all $A\in{\cal F}$.

The analysis of the solutions of (3.4) is rather tricky. It is
natural to begin with the translation - invariant solutions where
$h_x=h$ is constant for all $x\in V$. It is clear that a
Gibbs measure corresponding to this solution is translation-invariant.
This case has been investigated in Ref.\cite{GPW}.

In this case from (3.4) we virtue
$$
u={\theta_1^2\theta u^2+2\theta_1 u+
\theta \over \theta u^2+2\theta_1 u+\theta_1^2\theta } \eqno  (3.13)$$
where  $u=e^{2h}$.

\indent{\bf Proposition 3.2} \cite{GPW}.{\it  If $\theta_1
>\sqrt{3}$ and $\theta>\dsf{2\theta_1}{\theta_1^2-3}$ then  for
all pairs $(\theta,\theta_1)$ the equation (3.13) has three
positive solutions $u^*_{1}<u_{2}^{*}<u_{3}^*$, here $u^*_2=1$.
Otherwise the eq. (3.13) has a unique solution $u_*=1$.}\\

{\bf Remark.} The numbers $u^*_{1}$ and $u_{3}^*$ are the
solutions of the following equation
$$
u^2+(1+\a)u+1=0, \eqno(3.14)
$$
here $\a=\dsf{2\theta_1}{\theta}-\theta_1^2$. Hence $u^*_1u^*_3=1$
and if $\b\to\infty$ then $u^*_3\to\infty$ and $u^*_1\to 0$.

By $\mu_1,\mu_2,\mu_3$  we denote Gibbs measures corresponding to
these solutions.

Denote $u_x=\exp(2h_x), x\in V$. Then the functional equation
(3.4) is rewritten as
follows
$$
u_x=\frac{\theta_1^2\theta u_yu_z+\theta_1(u_y+u_z)+\theta}
{\theta u_yu_z+\theta_1(u_y+u_z)+\theta_1^2\theta}
\eqno  (3.15)
$$
here as before  $<y,x,z>$ are ternary neighboring vertices.

{\bf Proposition 3.3.} {\it Let  $\theta_1 >\sqrt{3}$,
$\theta>\dsf{2\theta_1}{\theta_1^2-3}$ and $u_x$ be a
solution of equation (3.15). Then $$ u_{1}^*\leq u_x\leq u_{3}^* \
\ \ \mbox{for any} \ \ x\in V. $$}

The Proof is similar to the proof of Proposition 4.3. of
\cite{MR}.

From this proposition we infer the following (See \cite{Ge})

{\bf Theorem 3.4.} {\it For the model (2.1) with parameters
$J_1>0$ and $J\in {\R}$ on the Cayley tree $\G^2$ the following
assertions hold}
\begin{enumerate}
   \item[(i)] {\it if $\theta_1 >\sqrt{3}$,
$\theta>\dsf{2\theta_1}{\theta_1^2-3}$ then the measures $\m_1$
and $\m_3$ are extreme;} \item[(ii)] {\it in the opposite case
there is a  Gibbs measure $\m_*(=\m_2)$
and it is extreme.}\\
\end{enumerate}

{\bf Remark.} This theorem specifies the result obtained in
\cite{GPW} as they has proved that a phase transition occurs if
and only if the above indicated conditions is satisfied, and the
extremity was open.  The formulated theorem answers that the found
Gibbs measures are extreme. Note that in the (i) case the measure
$\m_2$ need not to be extreme, some relevant information
concerning the extremity of $\m_2$ for the Ising model can be
found in \cite{BRZ1}.

It is clear from the construction of the Gibbs measures that the
measures $\m_1$ and $\m_3$ depend on parameter $\b$. Now we are
interested on the beheviour of these measures when $\b$ goes to
$\infty$.

Put
$$
\s_+=\{\s(x): \s(x)=1, \ \forall x\in\G^2\},
$$
$$
\s_-=\{\s(x): \s(x)=-1, \ \forall x\in\G^2\}.
$$

{\bf Theorem 3.5.} {\it Let  $\theta_1 >\sqrt{3}$ and
$\theta>\dsf{2\theta_1}{\theta_1^2-3}$, then
$$
\m_1\to \delta_{\s_-}, \ \ \ \m_3\to \delta_{\s_+}
\ \ \ \textrm{as} \ \ \b\to\infty,
$$
here $\delta_\s$ is a delta-measure concentrated on $\s$.}

{\bf Proof.} Consider the measure $\m_3$. This measure corresponds
to the function $h_x=h_3,$ $x\in V$, here $h_3>0$ (see Proposition
3.2). Let us first consider a case:
$$
\m_3(\s(x)=1)=\dsf{e^{h_3}}{e^{h_3}+e^{-h_3}}=\dsf{u^*_3}{u^*_3+1}\to
1 \ \ \textrm{as} \  \b\to\infty,
$$
since $u^*_3\to\infty$ as $\b\to\infty$, here $x\in V$. Let us
turn to the general case. From the condition imposed in the
Theorem we find that $J_1>0$. Now separately consider two cases.

{\tt First case.} Let $J>0$. Then from the form of Hamiltonian
(2.1) it is easy to check that $H(\s_n|_{V_n})\geq
H(\s_{+}|_{V_n})$ for all $\s\in \Omega$  and $n>0$. It follows
that \bea \m_3(\s_+|_{V_n})=\dsf{\exp\{-\b
H(\s_+|_{V_n})+h_3|W_n|\}}{\sum\limits_{\tilde\s_n\in\O_{V_n}}
\exp\{-\b H(\tilde\s_n)+h_3\sum\limits_{x\in W_n}\tilde\s(x)\}}=\nonumber \\
=\dsf{1}{1+\sum\limits_{\tilde\s_n\in\O_{V_n},\tilde \s_n\neq
\s_+|_{V_n}}
\dsf{\exp\{-\b H(\tilde\s_n)+h_3\sum\limits_{x\in W_n}\tilde\s(x)\}}{\exp\{-\b H(\s_+|_{V_n})+h_3|W_n|\}}}\geq\nonumber \\
\geq\dsf{1}{1+1/u^*_3}\to 1 \ \ \textrm{as} \
\b\to\infty.\nonumber \eea

The last inequality yields that $\m_3\to\delta_{\s_+}$.

{\tt Second case.} Let $J<0$. Let us introduce some notations.
$$
A(\s_n)=\sum_{>x,y<: x,y\in V_n}\s(x)\s(y), \ \ A=A(\s_+|_{V_n})
$$
$$
B(\s_n)=\sum_{<x,y>: x,y\in V_n}\s(x)\s(y), \ \ B=B(\s_+|_{V_n})
$$
$$
C(\s_n)=\sum_{x\in V_n}\s(x), \ \ C=C(\s_+|_{V_n}).
$$
Then it is easy to see that the following equality holds
$$
\m_3(\s_+|_{V_n})
=\dsf{1}{1+\sum\limits_{\tilde\s_n\in\O_{V_n},\tilde \s_n\neq
\s_+|_{V_n}}
\dsf{1}{e^{J\b(A-A(\tilde\s_n))}e^{J_1\b(B-B(\tilde\s_n))}e^{h_3(C-C(\tilde\s_n))}}}.
$$
We want to show that
$$
\sum_{\tilde\s_n\in\O_{V_n},\tilde \s_n\neq \s_+|_{V_n}}
\dsf{1}{e^{J\b(A-A(\tilde\s_n))}e^{J_1\b(B-B(\tilde\s_n))}e^{h_3(C-C(\tilde\s_n))}}\to
0 \ \textrm{as}
 \ \b\to\infty.
$$
It is enough to prove that
$$
\dsf{1}{e^{J\b(A-A(\tilde\s_n))}e^{J_1\b(B-B(\tilde\s_n))}e^{h_3(C-C(\tilde\s_n))}}\to
0 \ \ \textrm{as}
 \ \ \b\to\infty
$$
for all $\tilde\s_n\in\O_{V_n},\tilde \s_n\neq \s_+|_{V_n}$. We
rewrite the last sentence as follows
$$
\dsf{1}{e^{J\b(A-A(\tilde\s_n))}e^{J_1\b(B-B(\tilde\s_n))}e^{h_3(C-C(\tilde\s_n))}}=
\dsf{1}{\t^{(A-A(\tilde\s_n))/2}\t_1^{(B-B(\tilde\s_n))/2}(u^*_3)^{(C-C(\tilde\s_n))/2}}\leq
$$
$$
\leq\dsf{(\t_1^2-3)^{(A-A(\tilde\s_n))/2}}{\t_1^{(A-A(\tilde\s_n)+B-B(\tilde\s_n))/2}
(u^*_3)^{(C-C(\tilde\s_n))/2}}\leq
$$
$$
\leq\dsf{(\t_1^2-3)^{(A-A(\tilde\s_n))/2}}{\t_1^{(A-A(\tilde\s_n)+B-B(\tilde\s_n))/2}
u^*_3}.\eqno(3.16)
$$
here we have used the inequality $\t>\dsf{2\t_1}{\t_1^2-3}$.

Obviously, if $\b$ is large enough we have
$$
\dsf{(\t_1^2-3)^{(A-A(\tilde\s_n))}}{\t_1^{A-A(\tilde\s_n)+B-B(\tilde\s_n)}}\sim
\dsf{\t_1^{2(A-A(\tilde\s_n))}}{\t_1^{A-A(\tilde\s_n)+B-B(\tilde\s_n)}}=
$$
$$
=\dsf{\t_1^{B(\tilde\s_n)-A(\tilde\s_n)}}{\t_1^{B-A}}.
$$

If $B(\tilde\s_n)-A(\tilde\s_n)\leq B-A$ then the last relation
implies that
$\dsf{(\t_1^2-3)^{(A-A(\tilde\s_n))}}{\t_1^{A-A(\tilde\s_n)+B-B(\tilde\s_n)}}
$ is bounded, and hence from (3.16) we get the desired relation.

Now it remains to prove the following

{\bf Lemma 3.6.} {\it For every $n>0$ and $\s_n\in \O_{V_n}$ the
following inequality holds
$$
B(\s_n)-A(\s_n)\leq B-A.\eqno(3.17)
$$
}

{\bf Proof.} Denote ${\cal C}(\s_n)=\{x\in V_n : \s(x)=-1\}$.
Maximal connected components of ${\cal C}(\s_n)$ we will denote by
${\cal K}_1(\s_n),\cdots,{\cal K}_m(\s_n)$. For a connected
subset ${\cal K}$ of $V_n$ put
$$
\partial{\cal K}=\{x\in V_n\setminus{\cal K} : \ <x,y> \ \textrm{for some}
 \ y\in{\cal K}\}.
$$
$$\partial^2{\cal K}=\{x\in V_n\setminus{\cal K} : \ >x,y< \ \textrm{for some} \ y\in{\cal K}\}.
$$
From definition of $A(\s_n)$ and $B(\s_n)$ we get
$$
B(\s_n)=B-2\sum_j|\partial{\cal K}_j(\s_n)|,
$$
$$
A(\s_n)=A-2\sum_j|\partial^2{\cal K}_j(\s_n)\setminus \cup_{m\ne
j}{\cal K}_m(\s_n)|,
$$
here $|A|$ stands for a number of elements of a set $A$.

To prove (3.17) it enough to show that $|\partial^2{\cal K}|\leq
|\partial{\cal K}|$ for all connected subsets   ${\cal K}$ of
$V_n$. For each $x\in \partial^2 {\cal K}$ we can show a
$y=y(x)\in \partial {\cal K}$. Indeed, if $>x,t<, \ \ t\in {\cal
K}$ and $<x,z,t>$ then $y(x)=x $ if $z\in {\cal K}$ and $y(x)=z$
if
 $z\notin {\cal K}.$ It is clear $y(x)\in \partial {\cal K}.$ Now we will prove that
 if $x_1\ne x_2\in \partial^2 {\cal K}$ then $y(x_1)\ne y(x_2)\in \partial {\cal K}$.
 Let $<x_1,z_1,t_1>, <x_2,z_2,t_2>,$ where $t_1, t_2\in {\cal K}$. By definition of
 $y(x)$ we have $y(x_i)\in \{x_i,z_i\}, i=1,2$. So to prove $y(x_1)\ne
y(x_2)$ for $x_1\ne x_2$ it is enough to show $z_1\ne z_2$. Note
that in our case (i.e. $k=2$) if $x_1\ne x_2\in \partial^2 {\cal
K}$ then $d(x_1,x_2)\geq 3$, since $<x_i,z_i>, i=1,2$ and hence we
get $z_1\ne z_2$. Thus
 $|\partial^2{\cal K}|\leq |\partial{\cal K}|.$
 So Lemma is proved.

By similar argument the theorem can be proved for the measure
$\m_1$. Thus the theorem is proved.\\

From Theorem 3.5 we conclude that $\s_+$ and $\s_-$ are ground
states of the considered model.

{\bf Remark.} If in the condition  of Theorem 3.5, we put $J=0$
then the obtained result coincides with Theorem 2.3 of
\cite{BRZ2}. From proved theorem we infer that the measure $\m_2$ is not extreme while
$\b$ is large. For small $\b$ the extremity of $\m_2$ would be considered elsewhere. \\

\section{ A formula of the free energy}

Consider the partition function $Z_n(\b, h)$ of the state
$\m^h_\b$ (which corresponds to solution $h=\{h_x, x\in V\}$ of
the equation (3.4))
$$
 Z_n(\b,h)=\sum_{\tilde\s_n\in\Omega_{V_n}}\exp
\{-\b H(\tilde\s_n)+\sum_{x\in W_n}h_x\tilde\s(x)\}.
$$
The free energy is defined as
$$F(\b, h)=-\lim_{n\to \infty}{1\over 3\b\cdot 2^n}
\ln Z_n(\b, h).\eqno (4.1)$$
The goal of this section is to prove following

{\bf Theorem 4.1.}
\begin{enumerate}
   \item[(i)] {\it The free energy exists for all $h$, and is given by the
formula
$$F(\b, h)=-\lim_{\b\to\infty}\lim_{n\to\infty}{1\over 3\cdot 2^n}
\sum_{k=0}^n\sum_{x\in W_{n-k}}D_\b(J_1, J,h_y,h_z), \eqno(4.2)$$
where $y=y(x), z=z(x)$ are direct successors of $x$; \bea
D(J_1,J,h_y,h_z)=d_{J\b}(J_1\b+h_z)+d_{J\b}(-J_1\b+h_z)+ \nonumber
\\
\Delta_{J_1\b} (h_y+f(-J_1\b+h_z; \theta);h_y+f(J_1\b+h_z;
\theta));\nonumber \eea
$$
d_\b(x)={1\over 4}\ln[4\cosh(x-\b)\cosh(x+\b)];
$$
$$
\Delta_\b(x,y)={1\over
2}\ln[4\cosh(x-\b)\cosh(y+\b)];\eqno(4.3)$$
$$f(x,\theta)=\tanh^{-1}(\theta\tanh x), \ \ \theta=\tanh(J\b).$$}
\item[(ii)] {\it  For any solution $h=\{h_x, x\in V\}$ of (3.4)
$$F(\b, h)=F(\b, -h),$$ where $-h=\{-h_x, x\in V\}.$}
\end{enumerate}

{\bf Proof.} (i) We shall use the recursive equation (3.12):
$$Z_n=A_{n-1}Z_{n-1},$$
where $A_n=\prod\limits_{x\in W_n}a(x)$ and $a(x)=a(x, J_1,J,\b),
x\in V$
 is some function, which
we will define below. Using (3.10) we have \bea
a(x)=4\sqrt[4]{\cosh(J_1\b-J\b+h_z)\cosh(J_1\b+J\b+h_z)}\times\nonumber
\\
\sqrt[4]{\cosh(-J_1\b-J\b+h_z)\cosh(-J_1\b+J\b+h_z)}\times
\nonumber \\
\sqrt{\cosh(-J_1\b+h_y+f(-J_1\b+h_z;\theta))
\cosh(J_1\b+h_y+f(J_1\b+h_z;\theta))}.\nonumber \eea
 Thus, the
recursive equation (3.12) has the following form
$$Z_n(\b;h)=\exp\bigg(\sum_{x\in W_{n-1}}D_\b(J_1,J,h_y,h_z)\bigg)
Z_{n-1}(\b,h). \eqno(4.4)$$ This gives (4.2). Now we prove
existence of the RHS limit of (4.2). By proposition 3.4 we have
$h_x\in [{1\over 2}\ln u^*_1; {1\over 2}\ln u^*_3]$ consequently
functions from (4.3) and so $D$ is bounded i.e. $|D(J_1, J, h_y,
h_z)|\leq C_\b$ for all $h_y, h_z$. Hence we get
$${1\over 3\cdot 2^n}\sum_{k=l+1}^n\sum_{x\in W_{n-k}}D_\b(J_1, J, h_y,
h_z)\leq$$
$${C_\b\over  2^n}\sum_{k=l+1}^n 2^{n-k-1}
\leq C_\b\cdot 2^{-l}. \eqno (4.5)$$ Therefore, from (4.5) we get
the existence of the limit at RHS of (4.2).

(ii) Now, we shall prove that
$$F(\b, -h)=F(\b, h). \eqno(4.6)$$
It is easy to see that if $h=\{h_x, x\in V\}$ is a solution to
(3.4) then $-h=\{-h_x, x\in V\}$ also is solution to (3.4).
The equality (4.6) follows from the following
equality
$$D(J_1 ,J, - h_y, -h_z)=D(J_1, J, h_y, h_z),$$
which is a consequence of the following properties

1)\ \ $ d_\b(-x)=d_\b(x)$;

2)\ \  $f(-x; \theta)=-f(x;\theta);$

3)\ \  $\Delta_{\b}(-x,-y)=\Delta_{\b}(y,x). $

The theorem is proved.

The rest of the section is devoted to study of asymptotical
properties of the free energy $F(\b, h)$  as $\b\to \infty$ for
any constant $h$ i.e. $h_x=const$, $x\in V .$ In this setting
$F(\b, h)$ has the form:
$$F(\b, h)=D(J_1, J, h, h).\eqno(4.7)$$
Denote
$$ F(\infty)=\lim_{\b\to\infty}F(\b, h).\eqno (4.8)$$

By Proposition 3.2, we know that the equation (3.4) has exactly
three constant (translation -invariant) solutions:
$h=h(\b)=\{h_x={1\over 2}\ln u^*_1, x\in V\}$,
 $\{ h_x=0, x\in V\}$ and $\{h_x={1\over 2}\ln u^*_3, x\in V\} .$
Using (3.14) one can find that $h$ has the following asymptotic
$$h(\b)=M\b+o(\b^{-N}), \ \ N\geq 2, \ \  \mbox{as} \ \  \b\to\infty
\eqno(4.9)$$
where $M={1\over 2}\max\{2(J_1-J); 3J_1-J; 4J_1; J_1-J, 0\}.$
Using (4.9) it is easy to see that
$$d_{J\b}(J_1\b\pm h)={\b \over 4}\sum_{\e=\pm 1}|J_1+\e J\pm M|+o(\b^{-N}),$$
$$f(\pm J_1\b+M\b+o(\b^{-N});\theta)={\b\over 2}(|\pm J_1-J+M|-
|\pm J_1+J+M|)+o(\b^{-N});$$ here we have used the following
easily checking asymptotic
$$\ln \bigg(2\cosh(a\b+o(\b^{-N}))\bigg)=\b(|a|+o(\b^{-N})),$$ where $N\geq 2$ and
 $a\in \R.$
Consequently,
$$F(\b, h)=F(\infty)+ o(\b^{-N}), \eqno (4.10)$$
where
$$F(\infty)=\sum_{\delta=\pm 1}\left(\frac{1}{2}\bigg|\delta J_1+M-{1\over 2}
\sum_{\e=\pm 1}|\delta J_1+\b J+M|\e\bigg|+ {1\over 4}\sum_{\e=\pm
1}|J_1+\e J+\delta M|\right).$$

\section{Discussion of results}

It is known that to exact calculations in statistical mechanics
are paid attention by many of researchers, because those are
important not only for their own interest but also for some deeper
understanding of the critical properties of spin systems which are
not obtained form approximations. So, those are very useful for
testing the credibility and efficiency of any new method or
approximation before it is applied to more complicated spin
systems.  In this paper we have exactly solved an Ising model on a
Cayley tree, the Hamiltonian of which contains the
nearest-neighbor and competing interactions, namely,  we
calculated critical curve such that there is a phase transitions
above it, and a single Gibbs state is found elsewhere. It is found
ground states of the model. We also explicitly express the free
energy of the model under consideration, associated with the
translation invariant Gibbs measures, these enable us to find some
asymptotics of the free energy as $\b\to\infty$.\\

{\bf Acknowledgements.} The final part of the paper was done
within the scheme of Mathematical Fellowship (2004) at the Abdus
Salam ICTP. The authors thank ICTP for providing financial support
and all facilities. The first named author (F.M.) thanks Centro
Vito Volterra for kind hospitality, in particular Prof. L.Accardi
for discussions. The second named author (U.R.) also thanks
IMU/CDE-programme for travel support. The work partially supported
by Grants: $\Phi$-1.1.2 of CST of Uzbekistan  and
NATO-Reintegration Grant: FEL.RIG.980771.

\end{document}